\documentclass{webofc}
\usepackage[varg]{txfonts}   
\woctitle{Neutrinos, Torsion and Baryogenesis}
\begin{document}
\title{Neutrinos in the Early Universe, Kalb-Ramond Torsion and Matter-Antimatter Asymmetry}
%
%

\author{\underline{Nick E. Mavromatos}\inst{1,2}\fnsep\thanks{\email{Nikolaos.Mavromatos@kcl.ac.uk}}  \and
        Sarben Sarkar\inst{1}\fnsep\thanks{\email{Sarben.Sarkar@kcl.ac.uk}}
}

\institute{Theoretical Particle Physics and Cosmology Group, Department of Physics, King's College London, Strand, London WC2R 2LS, UK
\and
           Theory Division, Physics Department, CERN CH-1211 Geneva 23, Switzerland
          }

\abstract{%
The generation of a matter-antimatter asymmetry in the universe may be induced by the propagation of fermions in non-trivial, spherically asymmetric (and hence Lorentz violating) gravitational backgrounds. Such backgrounds may characterise the epoch of the early universe.  The key point in these models is that the background induces different dispersion relations, hence populations, between fermions and antifermions, and thus CPT Violation (CPTV) appears in thermal equilibrium. Species populations may freeze out leading to leptogenesis and baryogenesis. We consider here a string-inspired scenario, in which the CPTV is associated with a cosmological background with torsion provided by the  Kalb-Ramond (KR) antisymemtric tensor field of the string gravitational multiplet. In a four-dimensional space time this field is dual to a pseudoscalar ``axion-like'' field. The mixing of the KR field with an ordinary axion field can lead to the generation of a Majorana neutrino mass.
}
\maketitle
\section{Introduction}
\label{intro}

One of the most important issues of fundamental physics, relates to
an understanding of the magnitude of the observed baryon asymmetry $n_{B}-n_{\overline{B}}$ (where $B$ denotes baryon, $\overline{B}$ denotes antibaryon, $n_{B}$ is the number density of baryons and $n_{\overline{B}}$ the number density of antibaryons
in the universe). The universe is overwhelmingly made up of matter rather than anti-matter. According to the standard Big Bang theory, matter and antimatter have been created in equal amounts in the early
universe. However, the observed charge-parity (CP) violation in particle
physics~\cite{Christenson:1964fg}, prompted A. Sakharov~\cite{Sakharov:1967dj}
to conjecture that for baryon asymmetry in the universe (BAU) we need:
(i)  Baryon number violation to allow  for states with $\Delta B\neq 0$ starting from states with $\Delta B =0$ where $\Delta B$ is the change in baryon number. (ii)  If C or CP conjugate processes to a scattering process were allowed with the same amplitude  then baryon asymmetry would disappear. Hence C and CP need to be broken.
(iii)  Chemical equilibrium does not permit asymmetries. Consequently Sakharov required that chemical equilibrium does not hold during an epoch in the early universe. Hence non-equilibrium physics in the early universe together with
baryon number (B), charge (C) and charge-parity (CP) violating interactions/decays of anti-particles, may result in the observed BAU. In fact there are two types of non-equilibrium processes
in the early universe that can  produce this asymmetry: the first
type concerns processes generating asymmetries between leptons and
antileptons (\emph{leptogenesis}), while the second produces asymmetries
between baryons and antibaryons (\emph{baryogenesis}). The near complete
observed asymmetry today, is estimated in the Big-Bang theory~\cite{Gamow:1946eb}
to imply:
\begin{equation}
\Delta n(T\sim 1~{\rm GeV})=\frac{n_{B}-n_{\overline{B}}}{n_{B}+n_{\overline{B}}}\sim\frac{n_{B}-n_{\overline{B}}}{s}=(8.4-8.9)\times10^{-11}\label{basym}
\end{equation}
at the early stages of the expansion, e.g. for times $t<10^{-6}$~s
and temperatures $T>1$~GeV. In the above formula  $s$ denotes the
entropy density. Unfortunately, the observed CP violation within the
Standard Model (SM) of particle physics (found to be  $O(10^{-3})$ for the standard parameter $\epsilon$
in the neutral Kaon experiments~\cite{Christenson:1964fg}) induces
an asymmetry much less than that in (\ref{basym})~\cite{Kuzmin:1985mm}.
There are several ideas that go beyond the SM (\emph{e.g}.  grand unified
theories, supersymmetry, extra dimensional models \emph{etc.}) which involve
the decays of right-handed sterile neutrinos. For relevant important
works on this see \cite{Shaposhnikov:2009zz,Shaposhnikov:2006xi,Lindner:2010wr,Kusenko:2010ik,
Randall:1999ee,Merle:2011yv,Barry:2011wb}.
These ideas lead to extra sources for CP violation that could generate
the observed BAU. Some degree of fine tuning and somewhat \emph{ad hoc} assumptions are involved in such scenarios; so the quest for an understanding of the observed BAU still needs
further investigation. An example of fine tuning is provided by the choice of the hierarchy of the right-handed Majorana neutrino masses.
For instance, enhanced CP violation, necessary for BAU, can be achieved in models with three Majorana neutrinos, by assuming two of these neutrinos are
nearly degenerate in mass.

The requirement of non-equilibrium is on less firm ground~\cite{Carmona:2004xc}
than the other two requirements of Sakharov,
\emph{e.g.} if the non-equilibrium epoch occurred prior to inflation then its effects would be hugely diluted by inflation.
A basic assumption in the scenario of Sakharov is that \emph{CPT
symmetry}~\cite{cpttheorem}  (where $T$ denotes time reversal operation) holds in the very early universe.  \emph{CPT
symmetry} leads to the production
of matter and antimatter in equal amounts. Such \emph{CPT invariance}
is a cornerstone of all known \emph{local} effective \emph{relativistic}
field theories without gravity, and consequently of current particle-physics phenomenology. It should be noted that the necessity
of non-equilibrium processes in CPT invariant theories can be dropped if the
requirement of CPT invariance is relaxed~\cite{Bertolami:1996cq}. This violation of CPT (denoted by CPTV) is the result of a breakdown of Lorentz symmetry (which might happen at ultrahigh energies \cite{Mavromatos:2010pk}). For many
models with CPTV, in the time-line of the expanding universe, CPTV  generates
first lepton asymmetries (\textit{leptogenesis}); subsequently, through
sphaleron processes~\cite{PhysRevD.36.581} or baryon-lepton (B-L) number conserving processes
in Grand Unified Theories (GUT), the lepton asymmetry can be communicated
to the baryon sector to produce the observed BAU.

In order to obtain the observed BAU   CPTV in the early universe may obviate the need for fine tuning the decay widths of  extra sources of CP violation, such as sterile neutrinos and/or supersymmetry partners. Instead, one has to "tune" the background space-time, assuming a phase transition at an appropriate (high) temperature, after which the geometry of the universe assumes its canonical Robertson-Walker form. In this note we shall consider a simplified scenario~\cite{ems2013}:  the observed matter-antimatter asymmetry in the universe today is due to the coupling of right-handed Majorana neutrinos to a pseudoscalar background field  that originates from the Kalb-Ramond (KR) antisymmetric field of an ancestor string theory. The low energy limit of this ancestor string theory describes the observable universe. The oscillations of Majorana neutrinos between themselves and their antiparticles offer a microscopic realisation of chemical equilibrium processes which 
freeze out at a particular (high) temperature $T_D$ -the universe is assumed to undergo a phase transition such that the background KR field goes either to zero or to a very small value, compatible with the absence today of any observed CPTV effect. 
Such right-handed neutrinos characterise simple of the extensions of the Standard Model, termed neutrino-minimal-Standard-Model ($\nu$MSM)~\cite{Shaposhnikov:2009zz}, in the absence of supersymmetry or extra dimensions. $\nu$MSM  can provide candidates for dark matter. However, there are delicate issues associated with the realisation of the baryogenesis scenarios in this model, give that for the range of masses of the right-handed neutrinos employed in the model (two degenerate ones, with mass of order GeV, and a light one (dark matter), with mass of order O(10) keV); the baryogenesis is supposed to take place via coherent oscillations between the degenerate right-handed neutrinos.
Such coherent oscillations, though, may be destroyed in the high-temperature plasma of particles that characterises the early universe.

Our work provides a simple geometric scenario to avoid such dilemmas. We consider a  model such as the $\nu$MSM, in a KR background  which breaks Lorentz symmetry. The background couples to the right-handed neutrinos; a lepton asymmetry is induced by tuning the background. The crucial r\^ole of right-handed neutrinos for the realisation of our scenario~\cite{ems2013}, as sketched above, is compatible with the important r\^ole of the lightest of them as dark matter, envisaged in \cite{Shaposhnikov:2009zz, Boyarsky:2009ix}. Moreover 
in an era characterised by the apparent absence of supersymmetry signals in the large hadron collider (LHC)~\cite{mitsou2013}, 
the issue of the identification of the nature of the dark matter becomes even more pressing.

There is an additional significant r\^ole, for the KR axion field. Even if the background value of the field is zero in the present era, its quantum fluctuations, which survive today, may be responsible for giving the right-handed Majorana neutrinos their mass. This may happen 
through anomalous couplings of the KR field with the gravitational background and 
its mixing with an ordinary axion field, which couples via appropriate Yukawa couplings to 
the right-handed neutrinos~\cite{pilaftsis2012}. In this way, by an appropriate choice of the axion-neutrino Yukawa couplings, one may generate masses for the three right-handed neutrinos. Such masses lie in the range envisaged in $\nu$MSM~\cite{Shaposhnikov:2009zz}, so that the lightest of them (keV mass range) can play the r\^ole of a dark matter candidate.  The ordinary axions in this model may provide additional dark matter candidates. 

The structure of the talk, which is speculative, is the following: in the next section \ref{sec:2} we shall review some models where background geometries do not respect rotational symmetry, and so violate Lorentz symmetry (LIV). The background can induce CPTV matter-antimatter asymmetries in thermal equilibrium in the early universe. In section \ref{sec:3} we shall discuss our specific string-inspired model where the KR axion field plays the r\^ole of torsion. Torsion provides a LIV geometry and matter-antimatter asymmetry is generated.  We discuss right-handed neutrino-antineutrino oscillations of Pontercorvo type\cite{pontecorvo,Bilenky}; the oscillations violate \emph{both} CP and CPT. We also estimate the freeze-out temperature, which is the temperature at which the KR field switches off (or diminishes significantly) due to a phase transition of the string universe~\cite{ems2013}. 
In section \ref{sec:4} we discuss the r\^ole of the quantum fluctuations of the KR field in providing Majorana masses for the right-handed neutrinos.
Conclusions and an outlook appear in section \ref{sec:5}. 

\section{Lorentz-Violating Geometries and Matter-Antimatter Asymmetry in the Universe}
\label{sec:2}

We shall briefly review some existing models of CPTV induced asymmetry between matter and antimatter in the early universe. These existing models can be contrasted with our approach in this article. 

\subsection{CPTV Models with Particle-Antiparticle Mass Difference}

The simplest possibility~\cite{Dolgov:2009yk} for inducing CPTV
in the early universe is through particle-antiparticle mass differences
$m\ne\overline{m}$. These would affect the particle phase-space distribution
function $f(E,\mu)$,  $f(E,\mu)=[{\rm exp}(E-\mu)/T)\pm1]^{-1}~,\quad E^{2}=\vec{p}^{2}+m^{2}$, 
and antiparticle phase-space distribution function $ f(\overline{E},\bar{\mu})=[{\rm exp}(\bar{E}-\bar{\mu})/T)\pm1]^{-1}~,\quad\bar{E}^{2}=\vec{p}^{2}+\bar{m}^{2},$
with $\vec{p}$ being the $3-$momentum. (Our convention will be that
an overline over a quantity will refer to an antiparticle, $+$ will correspond to Fermi statistics (fermions), whereas $-$ will correspond to Bose statistics (bosons)).
Mass differences
between particles and antiparticles, $\overline{m}-m\neq0$, generate
a matter-antimatter asymmetry in the relevant number densities $n$ and $\overline{n}$, 
$n-\overline{n}=g_{d.o.f.}\int\frac{d^{3}p}{(2\pi)^{3}}[f(E,\mu)-f(\overline{E},\bar{\mu})], $
where $g_{d.o.f.}$ denotes the number of degrees of freedom of the
particle species under study. In the case of spontaneous Lorentz violation
\cite{Carroll:2005dj} there is a vector field $A_{\mu}$ with a non-zero
time-like expectation value which couples to a global current $J^{\mu}$
such as baryon number through an interaction lagrangian density
\begin{equation}
\mathcal{L}=\lambda A_{\mu}J^{\mu}.\label{LorVioln}
\end{equation}
This leads to $m\neq\bar{m}$ and $\mu\neq\bar{\mu}$. Alternatively,
following ~ \cite{Dolgov:2009yk} we can make the assumption that
the dominant contributions to baryon asymmetry come from quark-antiquark
mass differences, and that their masses ``run'' with the temperature i.e.
$m\sim gT$ (with $g$ the QCD coupling constant). One can provide
estimates for the induced baryon asymmetry on noting that the maximum
quark-antiquark mass difference is bounded by the current experimental
bound on the proton-antiproton mass difference, $\delta m_{p}(=|m_{p}-\overline{m}_{p}|)$,
known to be less than $\,2\cdot10^{-9}$ GeV. Taking $n_{\gamma}\sim0.24\, T^{3}$ (the
photon equilibrium density at temperature $T$) we have~\cite{Dolgov:2009yk}:
$ \beta_{T}=\frac{n_{B}}{n_{\gamma}}=8.4\times10^{-3}\,\frac{m_{u}\,\delta m_{u}+15m_{d}\,\delta m_{d}}{T^{2}}~,\quad\delta m_{q}=|m_{q}-{\overline{m}}_{q}|.$ Thus, $\beta_{T}$ is too small compared to the observed one. To reproduce
the observed $\beta_{T=0}\sim6\cdot10^{-10}$ one would need $\delta m_{q}(T=100~{\rm GeV})\sim10^{-5}-10^{-6}~{\rm GeV}\gg\delta m_{p}$,
which is somewhat unnatural.

However, active (\emph{light}) neutrino-antineutrino mass differences
alone may reproduce BAU; some phenomenological models in this direction
have been discussed in \cite{Barenboim:2001ac}, considering, for
instance, particle-antiparticle mass differences for active neutrinos
compatible with current oscillation data. This leads to the result
$n_{B}=n_{\nu}-n_{{\overline{\nu}}}\simeq\frac{\mu_{\nu}\, T^{2}}{6}$, 
 yielding $n_{B}/s\sim\frac{\mu_{\nu}}{T}\sim10^{-11}$ at $T\sim100$~GeV,
in agreement with the observed BAU. (Here $s$, $n_{\nu},\,\mathrm{and}\,\mu_{\nu}$
are the entropy density, neutrino density and chemical potential respectively.)

\subsection{CPTV-induced by Curvature effects in Background Geometry}\label{sec:bg}

In the literature
the r\^ole of gravity has been explicitly considered within a local effective
action framework which is essentially that of (\ref{LorVioln}). A
coupling to scalar curvature $R$ \cite{Davoudiasl:2004gf,Lambiase:2006md,Lambiase:2011by,Li:2004hh} through
a CP violating interaction Lagrangian $\mathcal{L}$:
$ \mathcal{L}=\frac{1}{M_{*}^{2}}\int d^{4}x\sqrt{-g}\left(\partial_{\mu} R \right)J^{\mu}$,
where $M_{*}$ is a cut-off in the effective field theory and $J^{\mu}$ could be the current associated with baryon (B) number. There is an implicit choice of sign in front of this interaction, which has been fixed so as to ensure matter dominance. It has been shown that \cite{Davoudiasl:2004gf}
$ \frac{n_{B-L}}{s}=\frac{\dot{R}}{M_{*}^{2} T_{d}},$ with
$T_{d}$ the freeze-out temperature for $B-L$ interactions. The idea then is that this asymmetry can be converted to baryon number asymmetry provided the $B+L$ violating (but B-L conserving) electroweak sphaleron interaction has not frozen out. To leading order in $M_{*}^{-2}$ we have $R=8\pi G (1-3w) \rho$ where $\rho$ is the energy density of matter and the equation of state is $p=w\rho$ where $p$ is pressure. For radiation $w=1/3$ and so in the radiation dominated era of the Friedmann-Robertson-Walker cosmology $R=0$. However $w$ is  precisely $1/3$ when $T^{\mu}_{\mu}=0$. In general $T^{\mu}_{\mu}\propto \beta(g)F^{\mu\nu}F_{\mu\nu}$  where $\beta(g)$ is the beta function of the running gauge coupling $g$ in a $SU(N_{c}$ gauge theory with $N_{c}$ colours. This allows $w\neq 1/3$. Further issues in this approach can be found in \cite{Davoudiasl:2004gf,Lambiase:2006md,Lambiase:2011by,Li:2004hh}.

 Another approach involves an axial vector current \cite{Debnath:2005wk,Mukhopadhyay:2005gb,Mukhopadhyay:2007vca,Sinha:2007uh}
instead of $J_{\mu}$. The scenario is based on the well known fact that fermions in curved
space-times exhibit a coupling of their spin to the curvature of the
background space-time.The Dirac Lagrangian density of a fermion can
be re-written as:
\begin{equation}\label{Bvector}
\mathcal{L}=\sqrt{-g}\,\overline{\psi}\left(i\gamma^{a}\partial_{a}-m+\gamma^{a}\gamma^{5}B_{a}\right)\psi~,\quad B^{d}=\epsilon^{abcd}e_{b\lambda}\left(\partial_{a}e_{\,\, c}^{\lambda}+\Gamma_{\nu\mu}^{\lambda}\, e_{\,\, c}^{\nu}\, e_{\,\, a}^{\mu}\right)~,
\end{equation}
in a standard notation, where $e_{\,\, a}^{\mu}$ are the vielbeins,
$\Gamma_{\,\alpha\beta}^{\mu}$ is the Christoffel connection and
Latin (Greek) letters denote tangent space (curved space-time) indices.
The space-time curvature background has, therefore, the effect of
inducing an ``axial'' background field $B_{a}$ which can be non-trivial
in certain anisotropic space-time geometries, such as Bianchi-type
cosmologies  or axisymmetric Kerr black holes~\cite{Debnath:2005wk,Mukhopadhyay:2005gb,Mukhopadhyay:2007vca,Sinha:2007uh}.
For an application to particle-antiparticle asymmetry it is necessary
for this axial field $B_{a}$ to be a constant in some local frame.
The existence of such a frame has not been demonstrated. As before
if it can be arranged that $B_{a}\neq0$ for $a=0$ then for constant
$B_{0}$ CPT is broken: the dispersion relation of neutrinos in such
backgrounds differs from that of antineutrinos. Explicitly, for the case of light-like $B_0 = |\vec B|$-background one has~\cite{dCMS}:
$(E \pm |\vec B|)^2 = (\vec p \pm \vec B )^2 + m^2 $, 
and for pure time-like B-backgrounds, of interest to us in the next section \ref{sec:3}~\cite{dCMS}, 
$E^2 = m^2 + (B_0 \pm |\vec p|)^2, $
where $m$ is the fermion mass and the $+$ ($-$) signs refer to particles (antiparticles) (in the case of Majorana neutrinos these are helicity states).
For small $m, B_0 << |\vec p|$ one may then obtain the (approximate) dispersion relations given in \cite{Mukhopadhyay:2005gb,Debnath:2005wk}
\begin{equation}\label{nunubardr}
E \simeq |\vec p| + \frac{m_{\rm eff}^2}{2 |\vec p|} + B_0 ~, \quad {\overline E} \simeq |\vec p| + \frac{m_{\rm eff}^2}{2 |\vec p|} - B_0~, \quad m_{\rm eff}^2 = m^2 + B_0^2 \ll |\vec p|~, 
\end{equation}
which we shall make use of in what follows. 

The relevant neutrino asymmetry emerges on following the same steps
used when there was
an explicit particle-antiparticle mass difference. As a consequence, for the pure-time like case considered above, and assuming 
a constant $B_0$, 
which will be of interest to us here, the following neutrino-antineutrino density difference is found from (\ref{nunubardr}):
$ \Delta n_{\nu}\equiv n_{\nu}-n_{\overline{\nu}}\sim g^{\star}\, T^{3}\left(\frac{B_{0}}{T}\right)$,
with $g^{\star}$ the number of degrees of freedom for the (relativistic)
neutrino. An excess of particles over antiparticles is predicted only
when $B_{0}>0$, which had to be assumed in the analysis of \cite{Debnath:2005wk,Mukhopadhyay:2005gb,Mukhopadhyay:2007vca,Sinha:2007uh};
we should note, however, that the sign of $B_{0}$ and its constancy have not been justified in
this phenomenological approach (The above considerations concern the dispersion relations for any fermion, not only neutrinos.
However, when one considers matter excitations from the vacuum, as relevant for leptogenesis, we need chiral fermions to get non trivial CPTV asymmetries in \emph{populations}
of particle and antiparticles, because $<\psi^\dagger \gamma^5 \psi> = - <\psi_L^\dagger \gamma^5 \psi_L> +  <\psi_R^\dagger \gamma^5 \psi_R> $.). At temperatures $T<T_{d}$, with $T_{d}$ the decoupling temperature
of the lepton-number violating processes, the ratio of the net Lepton
number $\Delta L$ (neutrino asymmetry) to entropy density (which
scales as $T^{3}$) remains constant,
\begin{equation}
\Delta L(T<T_{d})=\frac{\Delta n_{\nu}}{s}\sim\frac{B_{0}}{T_{d}}\label{dlbianchi}~.
\end{equation}
This  
implies a lepton asymmetry (leptogenesis) which, by tuning $B_0$ (for a given decoupling temperature $T_d$, that depends on the details of the underlying Lepton-number violating processes) can lead to a $\Delta L$ of the phenomenologically right order $\Delta L\sim10^{-10}$.
The latter can then be communicated
to the baryon sector to produce the observed BAU (baryogenesis)
by a B-L conserving symmetry in the context of either Grand Unified Theories
(GUT)~\cite{Debnath:2005wk}, or  sphaleron processes in the standard model. 

In the following section we shall discuss a case of a background where the constancy of $B_0$ in the Robertson-Walker cosmological frame 
is guaranteed by construction. This case is inspired by string theory. 

\section{Kalb-Ramond (KR) Torsion Background, Majorana Neutrinos and Baryogenesis}
\label{sec:3}

In this section we will discuss the case of a constant $B^0$ ``axial'' field that appears due to the interaction of the fermion spin with a string-theory background geometry with \emph{torsion}. This is a novel observation, which (as far as we are aware) was discussed for first time in \cite{mscptv}. 
In the presence of torsion the Christoffel symbol contains a part that is antisymmetric in its lower indices: $\Gamma^\lambda_{\,\,\,\,\mu\nu} \ne \Gamma^\lambda_{\,\,\,\,\nu\mu} $. Hence the last term of the right-hand side of the Eqn.(\ref{Bvector}) is \emph{not} zero.
Since the torsion term is of gravitational origin it couples universally to all fermion species. The effect of the coupling to neutrinos will be clarified below.

The massless gravitational multiplet in string theory contains the dilaton (spin 0, scalar), $\Phi$, the
graviton (spin 2, symmetric tensor), $g_{\mu\nu}$,  and the spin 1 antisymmetric tensor $B_{\mu\nu}$. The (Kalb-Ramond) field $B$ appears in the string effective action only through its totally antisymmetric field strength, $H_{\mu\nu\rho} = \partial_{\left[ \mu \right.} B_{\left.\nu\rho\right]}$, where  $[ \dots ]$ denotes antisymmetrization of the indices within the brackets. The calculation of  string amplitudes~\cite{sloan} shows that $H_{\mu\nu\rho}$ plays the role of \emph{torsion} in a generalised connection
$\overline{\Gamma}$:
\begin{equation}\label{generalised}
\overline{\Gamma}^\lambda_{\,\,\,\mu\nu} = \Gamma^\lambda_{\,\,\,\mu\nu} + e^{-2\Phi} H^\lambda_{\mu\nu} \equiv
\Gamma^\lambda_{\,\,\,\mu\nu} + T^\lambda_{\,\,\,\mu\nu}~.
\end{equation}
$\Gamma^\lambda_{\,\,\,\,\mu\nu} = \Gamma^\lambda_{\,\,\,\,\nu\mu} $ is the torsion-free Einstein-metric connection, and $T^\lambda_{\,\,\, \mu\nu} = - T^\lambda_{\,\,\,\nu\mu}$ is the \emph{torsion}.

In ref. \cite{aben} exact solutions to the conformal invariance conditions (to all orders in $\alpha^\prime$) of the low energy effective action of strings have been presented.  In four ``large'' (uncompactified)  dimensions of the string, the antisymmetric tensor field strength
can be written uniquely as
\begin{equation}\label{Hfield}
H_{\mu\nu\rho} = e^{2\Phi} \epsilon_{\mu\nu\rho\sigma} \partial^\sigma b (x)
\end{equation}
with $\epsilon_{0123} = \sqrt{g}$ and $\epsilon^{\mu\nu\rho\sigma} = |g|^{-1} \epsilon_{\mu\nu\rho\sigma}$, with $g$ the metric determinant.  The field
$b(x)$ is a ``pseudoscalar '' \emph{axion}-like field. The dilaton $\Phi$ and axion $b$ fields are fields that appear as Goldstone bosons of spontaneously broken scale symmetries of the string vacua, and so are exactly massless classically. In the effective string action such fields appear only through their derivatives.The exact solution of \cite{aben} in the \textit{string frame} requires that both dilaton and axion fields are linear in the target time $X^0$, $\Phi (X^0) \sim X^0$, $b(X^0) \sim X^0$. This solution will shift the minima of all fields in the effective action which couple to the dilaton and axion by a space-time independent amount.

In the ``physical''  \textit{Einstein frame }, relevant for cosmological observations, the temporal component of the
metric is normalised to $g_{00} =+1$ by an appropriate change of the time coordinate. In this setting,
the solution of \cite{aben} leads to a Friedmann-Robertson-Walker (FRW) metric, with scale factor $a(t) \sim t$, where $t$ is the FRW cosmic time. Moreover,  the dilaton field  $\Phi$ behaves as $- {\rm ln} t + \phi_0 $, with $\phi_0$ a constant, and the axion field $b(x)$ is  linear in $t$. There is an underlying world-sheet conformal field theory with central charge $c = 4 - 12 Q^2 - \frac{6}{n + 2} + c_I$ where $Q^2 (> 0 )$ is the central-charge deficit and  $c_I$ is the central charge associated with the world-sheet conformal field theory of the compact ``internal'' dimensions of the string model~\cite{aben}. The condition of cancellation of the world-sheet ghosts that appear because of the fixing of reparametrisation invariance of world-sheet co-ordinates requires that $c=26$. The solution for the axion field is
\begin{equation}\label{axion}
b(x) = \sqrt{2} e^{-\phi_0} \, \sqrt{Q^2} \,  \frac{M_s}{\sqrt{n}} t~,
\end{equation}
where $M_s$ is the string mass scale and $n$ is a positive integer, associated with the level of the Kac-Moody algebra of the underlying world-sheet conformal field theory. For non-zero $Q^2 $ there is an additional  dark energy term in  the effective target-space time action of the string~\cite{aben} of the form
$\int d^4 x \sqrt{-g} e^{2\Phi} (- 4Q^2)/\alpha^\prime $.
The linear axion field (\ref{axion}) \textit{remains} a non-trivial solution \emph{even} in the \emph{static} space-time limit with a constant dilaton field~\cite{aben}.  In such a case space time is an Einstein universe with positive cosmological constant and constant positive curvature proportional to
$6/(n+2)$.

For the solutions of \cite{aben}, the covariant torsion tensor  $e^{-2\Phi} H_{\mu\nu\rho} $ is \emph{constant}. (This follows from (\ref{generalised}) and (\ref{Hfield})
since the exponential  dilaton factors cancel out in the relevant expressions. ) Only the spatial components of the torsion are nonzero in this case,
\begin{equation}
T_{ijk} \sim \epsilon_{ijk} {\dot b} = \epsilon_{ijk} \sqrt{2 Q^2} e^{-\phi_0} \,  \frac{M_s}{\sqrt{n}}~,
\label{Hijk}
\end{equation}
where the overdot denotes derivative with respect to $t$.  As discussed in \cite{mscptv}, in the framework of the target-space effective theory, the relevant Lagrangian terms for fermions (to lowest order in $\alpha^\prime$) will be of the form (\ref{Bvector}), with the 
vector $B^0$ being associated with the spatial components of the constant torsion part $B^0 \sim \epsilon^{ijk} T_{ijk} $, where 
From (\ref{generalised}), (\ref{Hfield}) and (\ref{Bvector}), we also observe that only the temporal component $B^0$  of the $B^d$ vector is nonzero.  Note that the torsion-free gravitational part of the connection (for the FRW or flat case) yields a vanishing contribution to $B^0$.
From (\ref{Bvector}) and (\ref{Hijk}) then we obtain a constant $B^0$ of order
\begin{equation}\label{b0string}
B^0 \sim \sqrt{2 Q^2} e^{-\phi_0} \, \frac{M_s}{\sqrt{n}} ~ {\rm GeV} > 0.
\end{equation}
We follow the conventions of string theory for the sign of $B^0$ . From phenomenological considerations $M_s $ and  $g_s^2/4\pi$ are taken to be larger than O(10$^4$) GeV and  about $1/20$ respectively.

The particle-antiparticle asymmetry occurs already in thermal equilibrium, due to the background-induced difference in the dispersion relations between particles and antiparticles. Since the coupling of fermions to torsion is \emph{universal}, the axion background would also couple to quarks and charged leptons. However, it is the right-handed neutrinos that play a crucial r\^ole and induce a phenomenologically viable leptogenesis, and then baryogenesis through sphaleron processes in the standard model or other B-L conserving processes. 
This is due to the fact that, as argued in \cite{ems2013}, the right-handed Majorana neutrinos can oscillate  between themselves and their antiparticles, unlike the charged fermions of the standard model. Such $B_0$-background-induced neutrino-antineutrino oscillations, which have been envisaged initially by Pontercorvo~\cite{pontecorvo,Bilenky}, are induced by the mixing of neutrino and antineutrino states to produce mass eigenstates
due to the constant `environmental' field
$B^0$~\cite{Mukhopadhyay:2007vca,Sinha:2007uh}. To see this, we
consider the Lagrangian for Majorana neutrinos in the presence of $B_a$,
written in terms of two-component (Weyl) spinor fields  $\psi, \psi^c$ 
(since a generic four-component Majorana spinor $\Psi$ may be written in our notation as 
$\Psi = \begin{pmatrix} \psi^c_L \\ \psi_L \end{pmatrix}$, 
where from now on we omit the left-handed suffix $L$):
\begin{equation}\label{nulagr}
{\mathcal L}_{\nu} = \sqrt{-g} \Big[ \big({\psi^c}^\dagger \quad \psi^\dagger \big) \frac{i}{2} \gamma^0 \, \gamma^\mu \, D_\mu \begin{pmatrix} \psi^c \\ \psi \end{pmatrix} - \big({\psi^c}^\dagger \quad \psi^\dagger \big) \begin{pmatrix} -B_0 \quad -m \\ -m \quad B_0 \end{pmatrix} \, \begin{pmatrix} \psi^c \\ \psi \end{pmatrix} \, ,
\end{equation}
where $D_\mu$ is the gravitational covariant derivative with respect to the torsion-free spin connection, and 
we assume for brevity that the neutrino has only lepton-number-violating Majorana-type masses.
We note that the energy eigenstates are appropriate linear combinations of the states $|\psi \rangle $ and $|\psi^c \rangle$.  We observe from (\ref{nulagr}) that, in the presence of torsion,
there are non-trivial and unequal diagonal lepton-number-conserving
entries in the ``mass'' matrix $\mathcal{M}$ for $\psi$ and $\psi^c$: 
${\mathcal M} = \begin{pmatrix} - B_0 \quad - m \\ - m \quad B_0 \end{pmatrix}$. 
This matrix is hermitean, so can be diagonalised by a unitary matrix, 
leading to two-component mass eigenstates $| \chi_{i,j} \rangle$ that are mixtures of the 
states  $|\psi \rangle$ and $|\psi^c \rangle$ (and hence of the energy eigenstates):
\begin{eqnarray}\label{masseig}
| \chi_1 \rangle  & = & {\mathcal N}^{-1} \, \{ \Big( B_0 + \sqrt{B_0^2 + m^2} \Big)\, | \psi^c \rangle + m \, |\psi \rangle \} ~,  \nonumber \\
| \chi_2 \rangle  & = & {\mathcal N}^{-1} \, \{ - m \, |\psi^c \rangle + \Big( B_0 + \sqrt{B_0^2 + m^2} \Big)\, | \psi  \rangle  \} ~,  
\end{eqnarray}
where ${\mathcal N} \equiv \Big[ 2 \Big(B_0^2 + m^2 + B_0 \sqrt{B_0^2 + m^2} \Big)\Big]^{1/2}$, 
with eigenvalues  $m_{1,2} = \mp \sqrt{B_0^2 + m^2} $.

The above mixing can be expressed by writing the four-component neutrino spinor in terms of
$\psi$ and $\psi^c$ using an angle $\theta$~\cite{Sinha:2007uh}:
\begin{equation}\label{mixingangle}
\nu \; \equiv \; \begin{pmatrix} \chi_1 \\ \chi_2 \end{pmatrix} = \begin{pmatrix} {\rm cos}\, \theta  \quad {\rm sin}\, \theta \\ -{\rm sin}\, \theta
\quad  {\rm cos} \, \theta \end{pmatrix} \, \begin{pmatrix} \psi^c \\ \psi \end{pmatrix}~: \qquad {\rm tan} \, \theta \equiv \frac{m}{B_0 + \sqrt{B_0^2 + m^2}}~.
\end{equation}
It is readily seen that the four-component spinor $\nu$ is also Majorana, 
as it satisfies the Majorana condition $\nu^c = \nu$. We note that in the absence of torsion, $B_0 \to 0$,
the mixing angle between the two-component spinors $\psi$ and $\psi^c$ is maximal: $\theta = \pi/4$,
whereas it is non-maximal when $B_0 \ne 0$.

The mixing (\ref{masseig}) enables us to understand the difference between the densities of
fermions and antifermions mentioned earlier (\ref{dlbianchi}).
The expectation values of the number operators of $\chi_{i} , i=1,2$ in the basis $|\psi \rangle $ and $|\psi^c \rangle$  are given by:
$ N_{\chi_1}  =  < : \chi_1^\dagger \, \chi_1: > =  {\rm cos}^2 \theta \, < : {\psi^c}^\dagger \, \psi^c : >  + {\rm sin}^2 \theta \, < : {\psi }^\dagger \, \psi : > $,  $N_{\chi_2} =  < :\chi_2^\dagger \, \chi_2: > =  {\rm sin}^2 \theta \, < : {\psi^c}^\dagger \, \psi^c : >  + {\rm cos}^2 \theta \, < : {\psi }^\dagger \, \psi : > $, 
where cross-terms do not contribute. We observe that, for general $\theta \ne \pi/4 $, \emph{i.e}., $B_0 \ne 0$,
as seen in (\ref{mixingangle}), there is a difference between the populations of $\chi_{1}$ and $\chi_{2}$:
$ N_{\chi_1} - N_{\chi_2} = {\rm cos}\, 2\theta \Big( <n_{\psi^c} > -  <n_{\psi} > \Big)$, 
where $<n_{\psi}> = <: \psi^\dagger \, \psi :> \ne <n_{\psi^c}> = <: {\psi^c}^\dagger \, \psi^c :>$ are the corresponding number operators for the states $|\psi \rangle $ and $|\psi^c \rangle$.
 
This difference in the neutrino and antineutrino populations (\ref{dlbianchi})
is made possible by the presence of fermion-number-violating fermion-antifermion oscillations,
whose probability was calculated in~\cite{Sinha:2007uh}: 
\begin{equation}\label{probmix}
{\mathcal P}(t) = |\langle \nu_1 (t) | \nu_2(0) \rangle |^2 \propto {\rm sin}^2 \theta  \, {\rm sin}^2 \Big(\frac{E_\nu - E_{\nu^c}}{2} \, t \, \Big) = \frac{m^2}{B_0^2 + m^2} \, {\rm sin}^2 (B_0 \, t)~,
\end{equation}
where we used  (\ref{nunubardr}) with $\vec B = 0$, as in our specific background (\ref{b0string})
and the definition of the mixing angle (\ref{mixingangle}). 
The time evolution of the system is calculated by expressing  $|\psi \rangle$ and $|\psi^c\rangle$ as appropriate linear combinations of the eigenstates of the Hamiltonian. This determines the argument of the sinusoidal oscillation term
${\rm sin}^2 \Big(\frac{E_\nu - E_{\nu^c}}{2} \, t \, \Big)$. In the case of relativistic neutrinos moving close to the speed of light, the oscillation length obtained from (\ref{probmix}) is 
\begin{equation}\label{osclength}
L = \frac{\pi\, \hbar \, c }{|B_0|} = \frac{6.3 \times 10^{-14}\, {\rm GeV} }{B_0}~{\rm cm}.
\end{equation}
where we have reinstated $\hbar$ and $c$, and $B_0$ is measured in GeV.

For oscillations to be effective at any given epoch in the early Universe, 
this length has to be less than the size of the Hubble horizon.
We assume that a cosmological solution of the form discussed in~\cite{aben}, with a scale factor increasing linearly with time, is applicable some time after any earlier inflationary epoch. 
For a temperature $T_d \sim 10^9$ GeV, the relevant Hubble horizon size $\sim 10^{-12}$ cm. On
the other hand, we see from (\ref{dlbianchi}) that the correct order of magnitude for the lepton asymmetry 
$\sim 10^{-10}$ is obtained if
$B_0 \sim 10^{-1}$ GeV. For this value of B0, the oscillation length (\ref{osclength}) $10^{-13}$ cm, which is within the Hubble horizon size $10^{-12}$ cm.  
This implies that Majorana neutrino/antineutrino oscillations occur sufficiently rapidly to establish
chemical equilibrium and hence a lepton asymmetry. On the other hand, as already mentioned, 
charged leptons and quarks, although coupled to the KR torsion $H$, nevertheless do not exhibit such oscillations due to charge conservation. 

At the temperature $T_d \simeq 10^9 $ GeV the universe is assumed to undergo a \emph{phase transition}~\cite{ems2013} towards either a vanishing $B_0$ or at least a very small $B_0$ compatible with the current limits, $B_0 < 10^{-2}$ eV ,
of the relevant parameter of the standard model extension~\cite{Coleman,Bluhm,Kharlanov,Romalis}.
In this scenario for leptogenesis no fine tuning for the width of  the pertinent CP violating processes in the lepton sector is required, in contrast to the case of conventional leptogenesis~\cite{Pilaftsis:2003gt,Pilaftsis:2004xx,Pilaftsis:2005rv,Shaposhnikov:2009zz,Boyarsky:2009ix}. However, the presence of right-handed neutrinos 
was essential, and this is consistent with the need for explaining the smallness of the active neutrino masses through see-saw mechanisms, or the r\^ole of sterile neutrinos as dark matter~\cite{Shaposhnikov:2009zz,Boyarsky:2009ix}.)
The reader should note that the range of neutrino masses (Gev and keV) invoked in the latter works is consistent with the approximations leading to (\ref{osclength}).

\section{KR Torsion Fluctuations and Majorana Mass Generation}
\label{sec:4}

Before concluding we would like to discuss another interesting aspect of the KR torsion: the generation
of the masses of the right-handed Majorana neutrinos used above, e.g. in the range of GeV and keV as required in the 
$\nu MSM$ model~\cite{Shaposhnikov:2009zz}. 
So far we have discussed the r\^ole of background KR torsion. However, as we discussed above, at the temperature $T_d $ the universe may undergo a phase transition to a vanishing $B_0$. The quantum fluctuations of the torsion, however, survive. In this section we would like to make a suggestion~\cite{pilaftsis2012} that links these fluctuations to a mechanism for dynamical generation of (chirality changing) Majorana mass terms for neutrinos.

To discuss quantum aspects of torsion we first notice that the KR H-torsion is a totally antisymmetric type of torsion
coupled to fermions as (using for brevity differential form language): $S_\psi \ni -\frac{3}{4} \int S \wedge ^\star J^5 $
where $J_\mu^5 = {\overline \psi} \gamma_\mu \, \gamma^5  \psi $ is the axial fermion current. Here the fermions $\psi$ are generic and represent all sermonic excitations of the Standard Model plus right handed Majorana neutrinos. 
The totally antisymmetric part of the torsion $S = ^\star T$, that is $S_d = \frac{1}{3!} \epsilon^{abc}_{\,\,\,\, d} T_{abc}$,
where $T_{abc} $ is the contorsion which is proportional to $H_{abc} = \epsilon_{abcd} \partial^d b (x)$ in our case, with $b$ the KR axion field. Classically one has the Bianchi identity $d^\star S= 0$. To discuss quantum corrections~\cite{pilaftsis2012} we impose the constrain that quantum corrections should not 
affect this Bianchi identity, which allows for a definition of a conserved torsion charge $Q = \int ^\star  S$. 
Implementing this constraint via a delta function in the relevant path integral $\delta (d^\star S)$ leads to the introduction of a Lagrange multiplier field $b$ 
\begin{eqnarray}
 \label{qtorsion}
&&\hspace{-5mm} {\mathcal Z} \ni \int D \textbf{S} \, D b   \, \exp \Big[ i \int
    \frac{3}{4\kappa^2} \textbf{S} \wedge {}^\star\! \textbf{S} -
      \frac{3}{4} \textbf{S} \wedge {}^\star\! \textbf{J}^5  +
      \Big(\frac{3}{2\kappa^2}\Big)^{1/2} \, b \, d {}^\star\! \textbf{S}
      \Big]\nonumber \\  
&&\hspace{-5mm}=\!  \int D b  \, \exp\Big[ -i \int \frac{1}{2}
      \textbf{d} b\wedge {}^\star\! \textbf{d} b + \frac{1}{f_b}\textbf{d}b 
\wedge {}^\star\! \textbf{J}^5 + \frac{1}{2f_b^2}
    \textbf{J}^5\wedge\textbf{J}^5 \Big]\; ,\nonumber\\
\end{eqnarray}
where  $f_b = (3\kappa^2/8)^{-1/2} = \frac{M_P}{\sqrt{3\pi}}$ 
and  the  non-propagating   $\textbf{S}$  field  has  been  integrated
out. Here we have used the same notation $b$ for the Lagrange multiplier field as the background KR axion field.
This is for reasons of economy. The field $b$ in (\ref{qtorsion}) denotes quantum fluctuations of the KR axion,
and we assume a vanishing background for this field today. If one considers the quantum fluctuations about the background then the background terms are understood (but not explicitly given) in (\ref{qtorsion}). 
The reader  should notice that, as a  result of this integration,
the   corresponding   \emph{effective}   field   theory   contains   a
\emph{non-renormalizable} repulsive four-fermion axial-current-current
interaction. By partially integrating the term $d b \wedge ^\star J^5 $ and using the (one-loop exact) chiral anomaly equation  $\nabla_\mu J^{5\mu} \!=\! \frac{e^2}{8\pi^2} {F}^{\mu\nu}
  \widetilde{F}_{\mu\nu}  
- \frac{1}{192\pi^2} {R}^{\mu\nu\rho\sigma} \widetilde
{R}_{\mu\nu\rho\sigma} $, where $\mathbf{F}$ denotes field strength of gauge fields, and $R$ is the four-dimensional space time gravitational curvature, we obtain an effective ``axion-like'' coupling for the KR axion with the gauge sector 
$S_{\rm eff} \ni - \frac{e^2 }{8 \pi^2 \, f_b} \, \int b(x) F^{\mu\nu} \, {\widetilde F}_{\mu\nu} + \frac{1}{192\pi^2\, f_b} 
\int b(x) R^{\mu\nu\rho\sigma} \, {\widetilde R}_{\mu\nu\rho\sigma}$, 
where the $({\widetilde \dots })$ notation denotes dual tensors. The important point to notice is that the $b$ axion field is massless, unlike the ordinary axion field.

We notice at this stage, that for the case of the electromagnetic field, 
the  term $b F_{\mu\nu}  {\widetilde F}^{\mu\nu} $ becomes (up to total derivative terms)  a Chern-Simons (CS) form in four space-time dimensions
$\int b F_{\mu\nu}  {\widetilde F}^{\mu\nu} \propto  S_{\rm CS} = \int B_\mu A_\nu F_{\rho\sigma} \epsilon^{\mu\nu\rho\sigma} ~, \quad
B_\mu = \epsilon_{\mu \alpha \beta \gamma} H^{\alpha \beta \gamma} ~, \, H_{\alpha \beta \gamma} = \epsilon_{\alpha\beta\gamma \delta} \partial^\delta b(x)$.  Notice that $B_\mu$ is nothing but our axial vector coupled to the fermions in  (\ref{Bvector}), but here is not a background but a full fledged quantum field. In fact, when considering the coupling of charged fermions (\emph{e.g.} electrons or quarks) with the electromagnetic field $A_\mu$, the presence of such CS terms may affect the quantum photon propagator. This subject is still controversial, and we postpone a detailed discussion for a forthcoming publication~\cite{dCMS}.

For the purposes of the current work, we notice that, following ref.~\cite{pilaftsis2012}, we may couple (via appropriate Yukawa interactions of strength $y_a$ ) the (right-handed) Majorana fermions to an ordinary axion field, $a(x)$, which is allowed to mix (via the corresponding kinetic terms $\gamma \int \partial_\mu  b \, \partial^\mu  a$, with $|\gamma| < 1$) with the KR axion $b(x)$. It is convenient to diagonalize  the axion kinetic terms by redefining
the KR axion field $ b(x) \rightarrow {b^\prime}(x) \equiv b(x) + \gamma a(x) $ and canonically normalise the 
axion field $a$. The $b^\prime $ field decouples, then, leaving an effective axion-fermion action~\cite{pilaftsis2012}:

\begin{figure}[t]
 \centering
  \includegraphics[clip,width=0.40\textwidth,height=0.15\textheight]{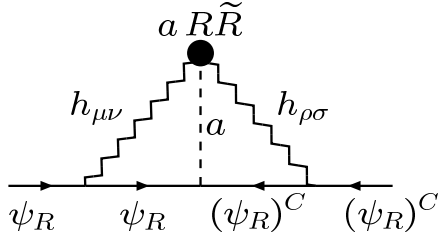} 
\caption{\it Feynman graph giving rise to anomalous fermion
  mass generation.  The black circle denotes the operator $a(x)\,
  R_{\mu\nu\lambda\rho}\widetilde{R}^{\mu\nu\lambda\rho}$ induced by
  torsion. Wavy lines are gravitons, dashed lines pertain to axion $a(x)$ propagators, while continuous lines denote Majorana spinors.}\label{fig:feyn}
\end{figure}
{\small
\begin{eqnarray} 
  \label{bacoupl3} 
\mathcal{S}_a \!\!=\!\! \int d^4 x
    \sqrt{-g} \, \Big[\frac{1}{2} (\partial_\mu a )^2 - \frac{\gamma
        a(x)}{192 \pi^2 f_b \sqrt{1 - \gamma^2}}
      {R}^{\mu\nu\rho\sigma} \widetilde{R}_{\mu\nu\rho\sigma} 
 - \frac{iy_a}{\sqrt{1 - \gamma^2}} \, 
a\, \Big( \overline{\psi}_R^{\ C} \psi_R - \overline{\psi}_R
\psi_R^{\ C}\Big) + \frac{1}{2f_b^2} J^5_\mu {J^5}^\mu
      \Big]\; .
\end{eqnarray}
\hspace{-1.5mm}}
The mechanism for  the anomalous Majorana mass generation  is shown in
Fig.~\ref{fig:feyn}.   We  may  now  estimate  the  two-loop  Majorana
neutrino mass in  quantum gravity with an effective  UV energy cut-off
$\Lambda$ by  adopting  the   effective   field-theory  framework   of
\cite{Donoghue}. 
This leads to a gravitationally induced Majorana mass~$M_R$:
$M_R \sim 
\frac{\sqrt{3}\, y_a\, \gamma\,  \kappa^5 \Lambda^6}{49152\sqrt{8}\,
\pi^4 (1 - \gamma^2 )}\; .$
In a UV
complete theory  such as  strings, $\Lambda$  and $M_P$  are related,
since $\Lambda$ is proportional to  $M_s$ and the latter is related to
$M_P$ (or  $\kappa$) via the strng coupling and the compactification volume.
Obviously, the generation of $M_R$ is highly model dependent.  Taking,
for example, the quantum gravity  scale to be $\Lambda = 10^{17}$~GeV,
we  find that  $M_R$ is  at the  TeV scale,  for $y_a  =  10^{-3}$ and
$\gamma =  0.1$. However, if we  take the quantum gravity  scale to be
close  to the  GUT scale,  i.e.~$\Lambda =  10^{16}$~GeV, we  obtain a
right-handed neutrino  mass $M_R \sim  16$~keV, for the choice  $y_a =
\gamma = 10^{-3}$.   This is in the preferred  ballpark region for the
sterile   neutrino    $\psi_R$   to    qualify   as   a    warm   dark
matter~\cite{Boyarsky:2009ix}. 

In a  string-theoretic framework, many  axions might exist  that could
mix  with each  other.  Such  a  mixing can  give rise  to reduced  UV
sensitivity of  the two-loop  graph shown in  Fig.~\ref{fig:feyn}.  To
make this point explicit, let  us therefore consider a scenario with a
number $n$ axion fields, $a_{1,2,\dots,n}$.  Of this collection of $n$
pseudoscalars, only $a_1$ has a  kinetic mixing term $\gamma$ with the
KR  axion  $b$  and  only   $a_n$  has  a  Yukawa  coupling  $y_a$  to
right-handed neutrinos  $\psi_R$.  The other  axions $a_{2,3,\dots,n}$
have a next-to-neighbour mixing pattern.  In such a model,  the anomalously  generated Majorana mass
may be estimated to be~\cite{pilaftsis2012}
$M_R \sim 
\frac{\sqrt{3}\, y_a\, \gamma\,  \kappa^5 \Lambda^{6-2n} 
(\delta M^2_a)^n}{49152\sqrt{8}\, 
\pi^4 (1 - \gamma^2 )}$, 
for $n \leq 3$, and thus independent of $\Lambda$ for $n=3$. 
Of  course, beyond the two  loops, $M_R$ will  depend on higher
powers of the energy  cut-off $\Lambda$, i.e.~$\Lambda^{n> 6}$, but if
$\kappa\Lambda \ll  1$, these higher-order effects are  expected to be
subdominant. In the above $n$-axion-mixing  scenarios, the anomalously
generated  Majorana mass  term  will only  depend  on the  mass-mixing
parameters $\delta M_a^2$ of the  axion fields and not on their masses
themselves, as long as $n \le 3$.

\section{Conclusions and Outlook}
\label{sec:5}

In this note we have discussed ways of obtaining leptogenesis/baryogenesis, which do not follow the Sakharov paradigm and involve
non-trivial background geometries of the early universe that violate Lorentz symmetry. As a specific example we considered a string-inspired theory involving anstisymmetric Kalb-Ramond (KR) tensor fields of spin 1, which in four space-time dimensions are equivalent to a pseudoscalar degree of freedom (the KR axion). The KR field provides the geometry with an appropriate totally antisymmetric torsion. 
The latter couples to all matter fermions both charged and neutral, but it is the coupling to right-handed Majorana 
neutrinos that plays a crucial r\^ole in providing microscopic processes of neutrino/antineutrino oscillations 
underlying the generation of matter-antimatter asymmetry in the lepton sector at high temperatures. The latter is then communicated to the baryon sector via the standard baryon-minus-lepton-number 
conserving sphaleron processes. The string universe is assumed to undergo a phase transition at a given temperature, at which the background KR axion field vanishes (or is diminished significantly, in agreement with the stringent bounds today on the Lorentz-symmetry-violating parameter of the standard model extension that corresponds to this background).

We have  also shown how quantum fluctuations of the KR torsion can generate an
effective (right-handed)  Majorana neutrino mass~$M_R$ 
at  two  loops  by  gravitational  interactions  that  involve  global
anomalies.  The  KR axion  $b$
couples to both matter and gravitation and radiation gauge fields.
In  perturbation  theory, this  axion  field $b$  derived from torsion has
derivative couplings, leading  to an axion shift symmetry:  $b \to b +
c$, where $c$ is an arbitrary constant.  If another axion field $a$ or
fields are  present in the theory,  the shift symmetry  may be broken,
giving rise to axion masses and chirality changing Yukawa couplings to
massless fermions, such as right-handed Majorana neutrinos $\psi_R$.

\section*{Acknowledgments}

We thank Marco de Cesare for discussions.  N.E.M. thanks the organisers of ICNFP 2013 Conference for the invitation to give a plenary talk.  The work of N.E.M. was supported in part by the London Centre for
Terauniverse Studies (LCTS), using funding from the European Research
Council via the Advanced Investigator Grant 267352, and 
by STFC UK under the research grant ST/J002798/1. 

 \bibliography{MSCPTVUniv2}

\end{document}